\documentclass{sig-alternate}
\usepackage{balance}  % to better equalize the last page
\usepackage{graphics} % for EPS, load graphicx instead 
\usepackage{txfonts}
\usepackage{pslatex}    % comment if you want LaTeX's default font
\usepackage[pdftex]{hyperref}
\usepackage{color}
\usepackage{textcomp}
\usepackage{booktabs}
\usepackage{ccicons}
\usepackage{subfigure}
\usepackage{enumitem}

\usepackage{pifont}

\usepackage{xspace}

\makeatletter
\def\url@leostyle{%
  \@ifundefined{selectfont}{\def\UrlFont{\sf}}{\def\UrlFont{\small\bf\ttfamily}}}
\makeatother
\def\pprw{8.5in}
\def\pprh{11in}

\usepackage{ifthen}
\usepackage[normalem]{ulem} % for \sout
\usepackage{xcolor}
\usepackage{amssymb}

\newboolean{showedits}
\setboolean{showedits}{true} % toggle to show or hide edits
\ifthenelse{\boolean{showedits}}
{
	\newcommand{\del}[1]{\textcolor{red}{\sout{#1}}} % please delete
}{
	\newcommand{\del}[1]{} % please delete
	
}

\newboolean{showcomments}
\setboolean{showcomments}{true}
\newcommand{\id}[1]{$-$Id: scgPaper.tex 32478 2010-04-29 09:11:32Z oscar $-$}

\ifthenelse{\boolean{showcomments}}
{\newcommand{\nbc}[3]{
 {\colorbox{#3}{\bfseries\sffamily\scriptsize\textcolor{white}{#1}}}
 {\textcolor{#3}{\sf\small$\blacktriangleright$\textit{#2}$\blacktriangleleft$}}}
 }
{\newcommand{\nbc}[3]{}
 \renewcommand{\del}[1]{} % please delete
 }

\definecolor{ibcolor}{rgb}{0.4,0.6,0.2}
\definecolor{pwcolor}{rgb}{0.8,0.3,0.2}

\usepackage{wasysym}

\usepackage{xcolor}

\definecolor{linkColor}{RGB}{6,125,233}
\hypersetup{%
  pdftitle={DoubleTake},
  pdfauthor={},
  pdfkeywords={},
  bookmarksnumbered,
  pdfstartview={FitH},
  colorlinks,
  citecolor=black,
  filecolor=black,
  linkcolor=black,
  urlcolor=linkColor,
  breaklinks=true,
}

\newcommand{\punt}[1]{}

\newcommand{\doubletake}{{\scshape DoubleTake}}
\newcommand{\DoubleTake}{{\scshape DoubleTake}}

\definecolor{lightgray}{rgb}{.9,.9,.9}
\definecolor{darkgray}{rgb}{.4,.4,.4}
\definecolor{purple}{rgb}{0.65, 0.12, 0.82}

\usepackage{comment} 
\usepackage{listings}

\lstdefinelanguage{c++threads}[]{c++}{
  morekeywords={pthread_create,pthread_join},
  keywordstyle=\color{blue}\bfseries,
  ndkeywords={class, export, boolean, throw, implements, import, this},
  ndkeywordstyle=\color{darkgray}\bfseries,
  identifierstyle=\color{black},
  sensitive=false,
  comment=[l]{//},
  morecomment=[s]{/*}{*/},
  commentstyle=\color{purple}\ttfamily,
  stringstyle=\color{red}\ttfamily,
  morestring=[b]',
  morestring=[b]"
}

\definecolor{Gray}{cmyk}{0,0,0,0.5}

\begin{document}

\CopyrightYear{2015}
%\copyrightdata{XXX-X-XXXXX-XXX-X/XX/XX}

\title{{\huge \bf \doubletake{}}: Fast and Precise Error Detection via Evidence-Based Dynamic Analysis}

%\authorinfo{\emph{authorship list removed for double-blind reviewing}}{}

\numberofauthors{3}

% Get the idea from http://tex.stackexchange.com/questions/9594/adding-more-than-one-author-with-different-affiliation

%\begin{comment}
%\author[1,2]{\rm Tongping~Liu}
%\author[1,3]{\rm Charlie~Curtsinger}
%\author[1]{\rm Emery~D.~Berger}
%\affil[1]{College of Information and Computer Sciences, University of Massachusetts Amherst}
%\affil[2]{Dept. of Computer Science, University of Texas at San Antonio}
%\affil[3]{Dept. of Computer Science, Grinnell College}
%\affil[ ]{\textit {Tongping.Liu@utsa.edu, curtsinger@cs.grinnell.edu, emery@cs.umass.edu}}
%\end{comment}

\author{
\alignauthor
Tongping~Liu \thanks{\footnotesize{This work was initiated and partially conducted while Liu and Curtsinger were PhD students at the University of Massachusetts Amherst.}}\\
\affaddr{Dept. of Computer Science} \\
\affaddr{University of Texas \\at San Antonio} \\
%\affaddr{One UTSA Circle} \\
\affaddr{San Antonio, TX 78249} \\
\email{Tongping.Liu@utsa.edu}
\alignauthor
Charlie~Curtsinger \footnotemark[1] \\
\affaddr{Dept. of Computer Science} \\
\affaddr{Grinnell College} \\
\affaddr{1116 8th Ave.} \\ 
\affaddr{Grinnell, IA 50112} \\
\email{curtsinger@cs.grinnell.edu}
\alignauthor
Emery~D.~Berger \\
\affaddr{College of Information and Computer Sciences} \\
\affaddr{University of Massachusetts Amherst} \\
\affaddr{Amherst, MA 01003} \\
\email{emery@cs.umass.edu}
}

\maketitle

\begin{comment}
\end{comment}

\begin{abstract}
Programs written in unsafe languages like C and C++ often suffer from
errors like buffer overflows, dangling pointers, and memory
leaks. Dynamic analysis tools like Valgrind can detect these errors,
but their overhead---primarily due to the cost of instrumenting every
memory read and write---makes them too heavyweight for use in deployed
applications and makes testing with them painfully slow. The result is
that much deployed software remains susceptible to these bugs, which
are notoriously difficult to track down.

This paper presents \emph{evidence-based dynamic analysis}, an
approach that enables lightweight analyses---under 5\% overhead for
these bugs---making it practical for the first time to perform these
analyses in deployed settings. The key insight of evidence-based
dynamic analysis is that for a class of errors, it is possible to
ensure that evidence that they happened at some point in the past
remains for later detection. Evidence-based dynamic analysis allows
execution to proceed at nearly full speed until the end of an epoch
(e.g., a heavyweight system call). It then examines program state to
check for evidence that an error occurred at some time during that
epoch. If so, it rolls back execution and re-executes the code with
instrumentation activated to pinpoint the error.

We present \doubletake{}, a prototype evidence-based dynamic analysis
framework. \doubletake{} is practical and easy to deploy, requiring
neither custom hardware, compiler, nor operating system support. We
demonstrate \doubletake{}'s generality and efficiency by building
dynamic analyses that find buffer overflows, memory use-after-free
errors, and memory leaks. Our evaluation shows that \doubletake{} is
efficient, imposing just 4\% overhead on average, making it the
fastest such system to date. It is also precise: \doubletake{}
pinpoints the location of these errors to the exact line and memory
addresses where they occur, providing valuable debugging information to
programmers.
\end{abstract}

%\category{D.1.3}{Programming Techniques}{Concurrent Programming--Parallel Programming}



%%%%%%%%%%%%%%%%%%%%%%%%%%%%%%%%%%%%%%%%%%%%%%%%%%%%%%%%%%%%%%%%%%%%%%%%%%%%%%%%%%%%%%%%%%%%%
%%%%%%%%%%%%%%%%%%%%%%%%%%%%%%%%%%%%%%%%%%%%%%%%%%%%%%%%%%%%%%%%%%%%%%%%%%%%%%%%%%%%%%%%%%%%%

\section{Introduction}
% Dynamic analysis
% Super-duper awesome

Dynamic analysis tools are widely used to find bugs in
applications. They are popular among programmers because of their
precision---for many analyses, they report no false positives---and
can pinpoint the exact location of errors, down to the individual line
of code. Perhaps the most prominent and widely used dynamic analysis
tool for C/C++ binaries is
Valgrind~\cite{overflow:valgrind}. Valgrind's most popular use case,
via its default tool, MemCheck, can find a wide range of memory
errors, including buffer overflows, use-after-free errors, and memory
leaks.

Unfortunately, these dynamic analysis tools often impose significant
performance overheads that make them prohibitive for use outside of
testing scenarios. An extreme example is the widely-used tool
Valgrind. Across the SPEC CPU2006 benchmark suite, Valgrind degrades
performance by almost 17$\times$ on average (geometric mean); its
overhead ranges from 4.5$\times$ and 42.8$\times$, making it often too
slow to use even for testing (see Table~\ref{table:valgrind}).

While faster dynamic analysis frameworks exist for finding particular
errors (leveraging compiler support to reduce overhead), they
sacrifice precision while continuing to impose substantial overhead
that would impede their use in deployed settings. The current
state-of-the-art, Google's AddressSanitizer, detects buffer overflows
and use-after-free errors, but slows applications by around
30\%~\cite{AddressSanitizer}. AddressSanitizer also identifies memory
leaks but only at the end of program execution, which is not useful
for servers or other long-lived applications.
%; precise memory leak detectors that identify leaked objects
%earlier remain far more expensive.

Because of their overhead, this class of dynamic analysis tools can
generally only be used during testing. However, they are limited by
definition to the executions that are tested prior to deployment. Even
exhaustive testing regimes will inevitably fail to uncover these
errors, which are notoriously difficult to debug.

This paper presents an approach called \emph{evidence-based dynamic
analysis} that is based on the following key insight: it is often
possible to discover evidence that an error occurred or plant markers
that ensure that such evidence exists. By combining evidence placement
with checkpointing and infrequent checking, we can run appplications
at nearly full speed in the common case (no errors). If we find an
error, we can use the checkpoint to roll back and re-execute the
program with instrumentation activated to pinpoint the exact cause of
the error.

Certain errors, including the ones we describe here, naturally exhibit
a \emph{monotonicity} property: when an error occurs, evidence that it
happened tends to remain or even grow so that it can be discovered at
a later point during execution. When this evidence is not naturally
occurring or not naturally monotonic, it can be forced to exhibit this
property by \emph{planting} evidence via what we call \emph{tripwires} to
ensure later detection. A canonical example of such a tripwire is a random
value, also known as a \emph{canary}, placed in unallocated space between
heap objects~\cite{StackGuard}. A corrupted canary is incontrovertible
evidence that a buffer overflow occurred at some time in the past.

This paper presents a prototype evidence-based dynamic analysis
framework called \doubletake{} that locates such errors with extremely
low overhead and no false positives. \doubletake{} checkpoints program
state and performs most of its error analyses only at epoch boundaries
(what we call irrevocable system calls) or when segfaults occur; these
occur relatively infrequently, amortizing \doubletake{}'s overhead.

If \doubletake{} finds evidence of an error at an epoch boundary or
after a segmentation violation, it re-executes the application from
the most recent checkpoint. During re-execution, \doubletake{} enables
instrumentation to let it precisely locate the source of the
error. For example, for buffer overflows, \doubletake{} sets hardware
watchpoints on the tripwire memory locations that were found to be
corrupted. During re-execution,
\doubletake{} pinpoints exactly the point where the buffer overflow
occurred.

We have implemented \doubletake{} as a drop-in library that
can be linked directly with the application, without the need to
modify code or even recompile the program. \doubletake{} works without
the need for custom hardware, compiler, or OS support.

Using \doubletake{} as a framework, we have built three different analyses that
attack three of the most salient problems for unsafe code: the buffer
overflow detector described above as well as a use-after-free detector
and memory leak detector. These analyses can all run concurrently. By
virtue of being evidence-based, they have a zero false positive rate,
precisely pinpoint the error location, and operate
with \emph{extremely} low overhead: for example, with \doubletake{},
buffer overflow analysis alone operates with just 3\% overhead on
average. When all three of these analyses are enabled, \doubletake{}'s
average overhead is under 5\%.

For all of the analyses we have implemented, \doubletake{} is the
fastest detector of these errors to date, providing compelling
evidence for the promise of evidence-based dynamic analyses. Its
overhead is already low enough to dramatically speed testing and often
low enough to enable the use of these formerly-prohibitive analyses in
deployed settings. This work thus promises to significantly extend the
reach of dynamic analyses.

\subsection*{Contributions}

The contributions of this paper are the following:

\begin{enumerate}

\item
It introduces \emph{evidence-based dynamic analysis}, a new analysis
technique that combines checkpointing with evidence gathering and
instrumented replay to enable precise error detection with extremely
low overhead.

\item
It presents \doubletake{}, a prototype framework that implements
evidence-based dynamic analyses for C/C++ programs: each of the
analyses we have built using \doubletake{} -- detecting buffer
overflows, use-after-frees, and memory leaks -- are the fastest
reported to date.

\end{enumerate}

\subsection*{Outline}
This paper first provides an overview of the basic operation
of \doubletake{} in
Section~\ref{sec:overview}. Section~\ref{sec:applications} details the
dynamic analyses we have built
using \doubletake{}. Section~\ref{sec:implementation} describes key
implementation details. Section~\ref{sec:evaluation} evaluates
\doubletake{}'s effectiveness, performance, and memory overhead, and
compares these to the state of the art.  Section~\ref{sec:discuss}
discusses limitations of evidence-based analysis and the detectors
we implement. Section~\ref{sec:relatedwork} describes
key related work and Section~\ref{sec:conclusion} concludes.

\begin{table}[!t]
\small
	\centering
	\begin{tabular}{l|r p{0.1em} l|r}
          \multicolumn{5}{c}{\textbf{\small Valgrind Execution Time Overhead}} \\
          \hline
		\textbf{Benchmark} & \textbf{Overhead} & & \textbf{Benchmark} & \textbf{Overhead} \\
		\cline{1-2} \cline{4-5}
		400.perlbench	& 20.5$\times$	& & 458.sjeng	& 20.3$\times$	\\
		401.bzip2		& 16.8$\times$	& & 471.omnetpp	& 13.9$\times$	\\
		403.gcc			& 18.7$\times$	& & 473.astar	& 11.9$\times$	\\
		429.mcf			& 4.5$\times$ 	& & 433.milc		& 11.0$\times$	\\
		445.gobmk		& 28.9$\times$	& & 444.namd		& 24.9$\times$	\\
		456.hmmer		& 13.8$\times$	& & 450.dealII	& 42.8$\times$	\\
	\end{tabular}
	\caption{Valgrind's execution time overhead across the SPEC benchmark suite. Valgrind imposes on average 17$\times$ overhead (geometric mean), making it prohibitively high for use in deployment and quite expensive even for testing purposes.\label{table:valgrind}}
\end{table}

\section{Overview}
\label{sec:overview}

\begin{figure}[!t]
\begin{center}
\includegraphics[width=3.3in]{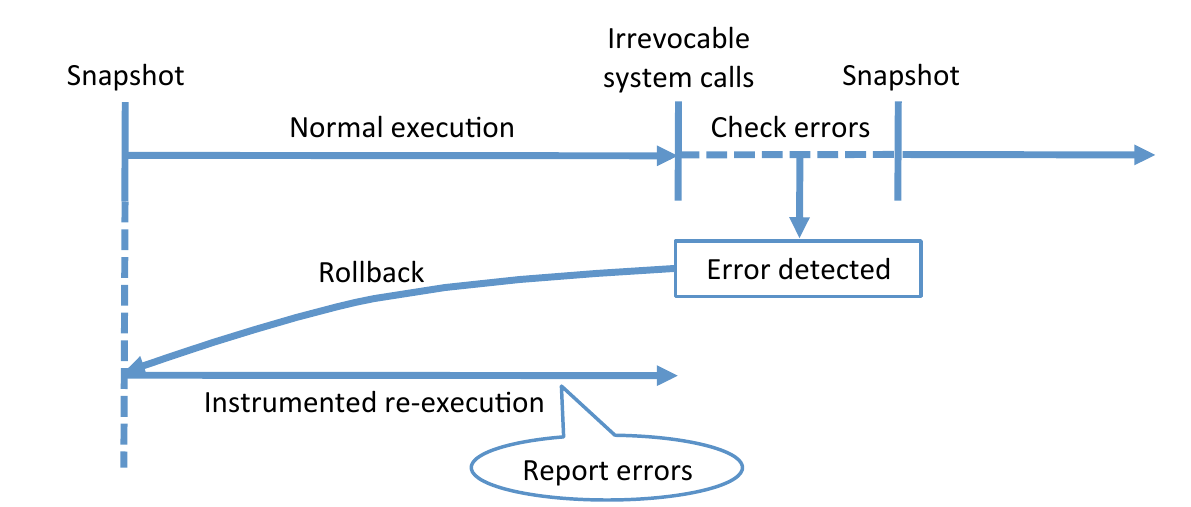}
\end{center}
\caption{
Overview of \doubletake{} in action: execution is divided into epochs
at the boundary of irrevocable system calls. Each epoch begins by
taking a snapshot of program state. Execution runs at nearly
full-speed during epochs. Evidence-based analysis takes place once an
epoch ends, replaying execution from the previous snapshot until it
pinpoints the exact location where the error is
introduced. Relatively-long epochs amortize the cost of snapshots and
analysis, keeping overhead low.
\label{fig:overview}}
\end{figure}

\doubletake{} is an efficient dynamic analysis framework for a class
of errors that exhibit or can be forced to exhibit
a \emph{monotonicity} property: evidence of the error is persistent
and can be gathered after-the-fact. With \doubletake{}, program
execution is divided into epochs, during which execution proceeds at
full speed (Figure~\ref{fig:overview}). At the beginning of an epoch,
\doubletake{} checkpoints program state. Epochs end only when the
application issues an \emph{irrevocable} system call (e.g., a socket
read); most system calls are not irrevocable (see \ref{sec:syscalls}
for full details). Once an epoch ends, \doubletake{} checks the
program state for evidence of memory errors. Because epochs are
relatively long-lived, the cost of checkpointing and error analysis is
amortized over program execution. If \doubletake{} finds an error, it
re-executes code executed from the previous epoch with additional
instrumentation to pinpoint the exact cause of the error.

To demonstrate \doubletake{}'s effectiveness, we have implemented
detection tools for three of the most important classes of errors in C
and C++ code: heap buffer overflows, use-after-free errors, and memory
leaks (Section~\ref{sec:applications} describes these in detail).  All
detection tools share the following core infrastructure
that \doubletake{} provides.

\subsection{Efficient Recording}

At the beginning of every epoch, \doubletake{} saves a snapshot of
program registers and all writable memory. An epoch ends when the
program attempts to issue an irrevocable system call, but most system
calls do not end the current epoch. \doubletake{} also records a small
amount of system state at the beginning of each epoch (e.g., file
offsets), which lets it unroll the effect of system calls that modify
this state when re-execution is required.

During execution, \doubletake{} manages various types of system calls
in an effort to reduce the number of epochs, which
Section~\ref{sec:implementation/normalexecution} discusses. In
practice, \doubletake{} limits the number of epoch boundaries,
amortizing the cost of program state checks. The kind of checks
employed depend on the particular dynamic analysis being performed;
Section~\ref{sec:applications} describes the details of the analyses
we have built on top of \doubletake{}.

\subsection{Lightweight Replay}

When program state checks indicate that an error occurred during the
current epoch, \doubletake{} replays execution from the last epoch to
pinpoint the error's root cause. \doubletake{} ensures that all
program-visible state, including system call results and memory
allocations and deallocations, is identical to the original
run. During replay, \doubletake{} returns cached return values for
most system calls, with special handling for some
cases. Section~\ref{sec:implementation} describes in detail how \doubletake{}
records and re-executes system calls.

%Note that \doubletake{}'s replay mechanism explicitly \emph{does not} guarantee deterministic replay in the face of race conditions. If it fails to detect an error on the second execution, that means that the error in the first execution was due to a race. \doubletake{} can then either continue execution, thus tolerating the race condition, or re-trigger the race by using schedule-perturbation techniques.

\subsection{Deterministic Memory Management and Tripwire Support}
\label{sec:heap}
One key challenge to using replay to find the exact location of errors is
that we cannot rely on the default system-supplied heap allocator. The reason for this is that it does not provide a replayable
sequence of addresses. The default heap grows on demand by
invoking \texttt{mmap} (or a similar call on other operating systems)
to obtain memory from the system. However, because of address-space
layout randomization, now implemented on all modern operating
systems to increase security, \texttt{mmap} almost always returns
different addresses when invoked. This effect means that heap
addresses in a replayed execution would likely differ from the
original.

%Another problem to use the default heap is that \doubletake{} has to identify the locations of  cannot know the range of the default heap beforehand. 

\doubletake{} therefore replaces the default heap allocator with a
heap built with the {\scshape Heap Layers}
framework~\cite{heaplayers}. In addition to providing repeatable
sequences of addresses, \doubletake{}'s heap provides a number
of other useful features that improve \doubletake{}'s efficiency and
simplify building analyses using it:

\begin{itemize}

\item \textbf{Efficiency via large chunk allocation.}
The \doubletake{} heap obtains memory from the operating system in large
chunks and satisfies all memory allocations from it, reducing the
number of system calls that \doubletake{} must track and thus lowering
its overhead.

\item \textbf{Simplified tripwire installation.}
\doubletake{}'s  heap also makes the process of implanting tripwires
easier. For example, detection tools can easily interpose on heap operations
to alter memory allocation requests or defer the reuse of freed
memory, and can mark the status of each object in metadata (e.g., via
a dedicated object header that the heap provides for this purpose).

\item \textbf{Efficient tripwire checking.}
Finally, \doubletake{}'s heap makes tripwire checking far more
efficient. It maintains a shadow bitmap to identify the locations and
status of heap canaries, which allows it to use vectorized bit
operations to perform efficient checking at the end of each epoch.

\end{itemize}

\noindent
Section~\ref{sec:heapallocator} presents full details
of \doubletake{}'s heap implementation.

\subsection{Pinpointing Error Locations}

During replay, \doubletake{} lets detection tools set \emph{hardware
watchpoints} during re-execution to pinpoint error locations (i.e., on
an overwritten canary). Modern architectures make available a small
number of watchpoints (four on x86). Each watchpoint can be configured
to pause program execution when a specific byte or word of memory is
accessed. While watchpoints are primarily used by
debuggers, \doubletake{} uses them to speed error location during
re-execution.

\doubletake{}'s watchpoints are
particularly useful in combination with heap canaries. For example,
during re-execution, \doubletake{}'s buffer overflow and
use-after-free detectors place a watchpoint at the location of the
overwritten canary to trap the instruction(s) responsible for the
error.

\section{Analyses}
\label{sec:applications}

To demonstrate \doubletake{}'s generality and efficiency, we implement
a range of error-detection tools as evidence-based dynamic
analyses. In particular, we implement the following three detection
tools with \doubletake{}:

\begin{itemize}
\item \textbf{Heap buffer overflow detection} ($\S$\ref{sec:applications/overflow}): when an application writes outside the bounds of an allocated object,
\item \textbf{Use-after-free detection} ($\S$\ref{sec:applications/useafterfree}): when an application writes to freed memory (i.e., through a \emph{dangling pointer}), and
\item \textbf{Memory leak detection} ($\S$\ref{sec:applications/leak}): when a heap object becomes inaccessible but has not been explicitly freed.
\end{itemize}

For each of these tools, we describe the evidence that \doubletake{}
observes or places to detect these errors, and how re-execution and error isolation
proceeds once an error is detected. Note that because these analyses
are orthogonal, they can all be used simultaneously.

\begin{figure}[!t]
\begin{center}
\includegraphics[width=3.3in]{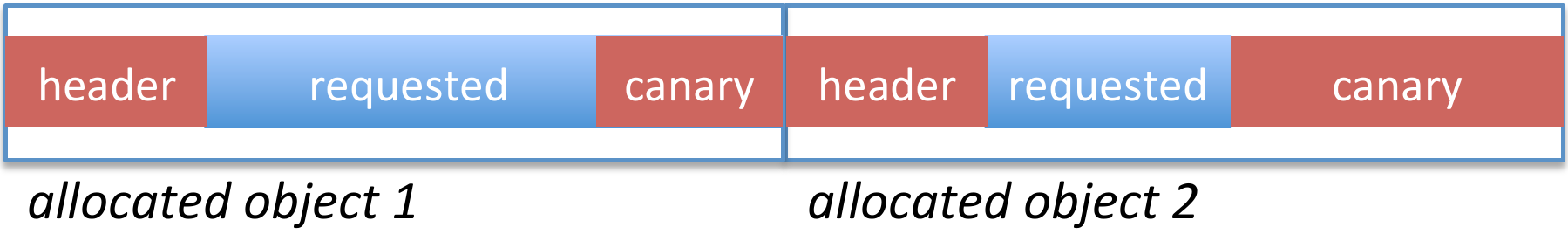}
\end{center}
\caption{Heap organization used to provide evidence of buffer overflow errors. Object headers and unrequested space within allocated objects are filled with canaries; a corrupted canary indicates an overflow occurred.
\label{fig:buffer-overflow}}
\end{figure}

\begin{figure}[!t]
\vspace{10pt}
\begin{center}
\includegraphics[width=3.3in]{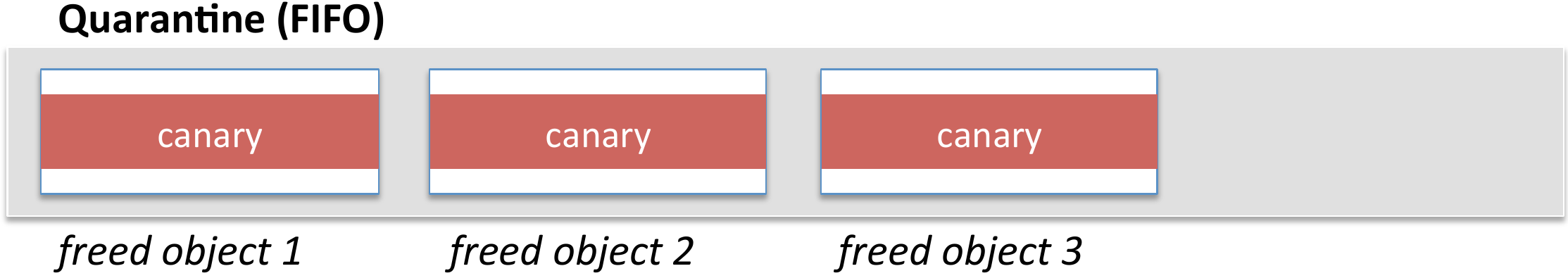}
\end{center}
\caption{Evidence-based detection of dangling pointer (use-after-free) errors. Freed objects are deferred in a quarantine in FIFO order and filled with canaries. A corrupted canary indicates that a write was performed after an object was freed. 
\label{fig:dangling-pointer}}
\end{figure}

\subsection{Heap Buffer Overflow Detection}
\label{sec:applications/overflow}
% Describe the basic setup, what evidence of an error will look like, and what the tool can report
Heap buffer overflows occur when programs write outside the bounds of an allocated object. 
\doubletake{} reports an error when it discovers that a canary value has been overwritten. When it finds an overwritten canary, the detector places watchpoints during re-execution to identify the instruction responsible for the overflow.

\subsubsection*{Evidence-Based Error Detection}
% Describe the implementation of detection

Figure~\ref{fig:buffer-overflow} presents an overview of the approach used to locate buffer overflows. Our buffer overflow detector places canaries between heap objects so that an overflow from one object into an adjacent one can be detected.

In addition, the overflow detector fills any remaining empty space
inside allocated objects with canaries; \doubletake{}'s allocator
rounds all object size requests up to the nearest power of two. This
approach lets \doubletake{} identify small overflows that would
otherwise be missed because they did not actually go beyond the
object's allocated space.

At memory deallocation time (calls to \texttt{free} or \texttt{delete}), \doubletake{} checks for buffer overflows in objects whose requested size is less than a power of two. It defers the checking of power-of-two sized objects to the end of the current epoch. 

At the end of each epoch, \doubletake{} checks whether any canaries have been overwritten (including those for exact power-of-two requests). If it finds any overwritten canaries, it has incontrovertible evidence that a buffer overflow has occurred. \doubletake{} then triggers a re-execution to locate the exact point in the program when the overflow happened.

\subsubsection*{Re-Execution and Error Isolation}
% Describe the procedure for re-executing and reporting additional details

\doubletake{} installs a watchpoint at the address of the corrupted canary before re-execution. When the program is re-executed, any instruction that writes to this address will trigger the watchpoint. The operating system will deliver a \texttt{SIGTRAP} signal to \doubletake{} before the instruction is executed. By handling this signal, \doubletake{} reports the complete call stack of the trapped instruction by invoking the \texttt{backtrace} function.

%%%%%%%%%%%%%%%%%%%%%%%%%%%%%

\subsection{Use-After-Free Detection}
% Describe the basic setup, what evidence of an error will look like, and what the tool can report
\label{sec:applications/useafterfree}
 
Use-after-free or dangling pointer overflow errors occur when an application continues to access memory through pointers that have been passed to \texttt{free()} or \texttt{delete}. Writes to freed memory can overwrite the contents of other live objects, leading to unexpected program behavior. Like the buffer overflow detector, our use-after-free detector uses canaries to detect writes to freed memory. When a use-after-free error is detected, \doubletake{} reports the allocation and deallocation sites of the object, and all instruction(s) that wrote to the object after it was freed.

\subsubsection*{Evidence-Based Error Detection}
% Describe the implementation of detection

Figure~\ref{fig:dangling-pointer} illustrates how we detect use-after-free errors using \doubletake{}. Our use-after-free detector delays the re-allocation of freed memory. We adopt the approach used by AddressSanitizer of maintaining a FIFO quarantine list~\cite{AddressSanitizer}. In our implementation, objects are released from the quarantine list when the total size of quarantined objects exceeds 16 megabytes, or when there are more than 1,024 quarantined objects. (Note that all thresholds used by the detector are easily configurable.)

The detector overwrites the first 128 bytes of all objects in the quarantine list  (which have all been freed by the program) with canary values. This threshold strikes a compromise between error detection and efficiency. We have found empirically that filling larger objects with canaries (i.e., going beyond 128 bytes to the full size of allocated objects) introduces substantial overhead during normal execution, but is unlikely to catch any additional errors. This is because large objects often consist of a header followed by a buffer. A prematurely reused object is likely to have its prologue scrambled by a constructor, while the remainder of the object (the buffer contents) may remain unmodified for a long time. 

Before an object can be returned to the program heap, \doubletake{}
verifies that no canaries have been overwritten. It also checks all
canaries in the entire heap at epoch boundaries. In either case, if a
canary has been overwritten, the detector knows that a use-after-free
error has occurred. It then immediately triggers re-execution to
identify the cause of this error.

\subsubsection*{Re-Execution and Error Isolation}
% Describe the procedure for re-executing and reporting additional details

During re-execution, the use-after-free detector interposes on \texttt{malloc} and \texttt{free} calls to find the allocation and deallocation sites of the overwritten object. The detector records a call stack for both sites using the \texttt{backtrace} function. The detector also installs a watchpoint at the address of the overwritten canary. As with buffer overflow detection, any writes to the watched address will generate a \texttt{SIGTRAP} signal. When this signal is triggered, the detector reports information about the object's allocation and deallocation sites, as well as call stack and line number information for the instructions responsible for the use-after-free error.

%%%%%%%%%%%%%%%%%%%%%%%%%%%%%

\subsection{Memory Leak Detection}
\label{sec:applications/leak}
% Describe the basic setup, what evidence of an error will look like, and what the tool can report
Heap memory is leaked when it becomes inaccessible without being freed. Memory leaks can significantly degrade program performance due to an increased memory footprint. Our leak detector identifies possible unreachable allocated objects at the end of each epoch. Allocation sites can help users fix memory leaks, but collecting this information for all\texttt{malloc} calls in normal execution would unnecessarily slow down the program  for the common case (no memory leaks). Instead, \doubletake{} only records the allocation sites of leaked memory during re-execution, and adds no overhead for normal execution.

\subsubsection*{Evidence-Based Error Detection}
% Describe the implementation of detection
Unlike the previously-described detectors, memory leak detection does not need tripwires. Instead, the evidence of a memory leak is latent in the heap organization itself.

Our detector finds memory leaks using the same marking approach as conservative garbage collection~\cite{Wilson:1992:UGC:645648.664824}. The marking phase performs a breadth-first scan of reachable memory using a work queue. Initially, all values in registers, globals, and the stack that look like pointers are added to the work queue. Any eight-byte aligned value that falls within the range of allocated heap memory is treated as a pointer.

At each step in the scan, the detector takes the first item off the work queue. Using the heap metadata located before each object, the detector finds the bounds of each object. Each object has a header containing a \emph{marked} bit and an \emph{allocated} bit. If the \emph{marked} bit is set, this object has already been visited. The detector then removes this object and moves on to the next item in the queue. If the object is allocated but not yet marked, the detector marks it as reachable by setting the \emph{marked} bit and adds all pointer values within the object's bounds to the work queue. Once the work queue is empty, \doubletake{} ends its scan.

\doubletake{} then traverses the entire heap to find any leaked objects: these are allocated but unmarked (unreachable). If it finds memory leaks, re-execution begins. Note that using this approach, our detector can also find potential dangling pointers (that is, reachable freed objects). This option is disabled by default because, unlike other applications, potential dangling pointer detection could produce false positives.

\subsubsection*{Re-Execution and Error Isolation}
% Describe the procedure for re-executing and reporting additional details

During re-execution, the leak detector checks the results of each \texttt{malloc} call. When the allocation of a leaked object is found, the detector records the call stack using the \texttt{backtrace} function. At the end of the epoch re-execution, the detector reports the last call stack for each leaked object since the last site is responsible for the memory leak. 

\section{Implementation Details}
\label{sec:implementation}

\doubletake{} is implemented as a library for Linux applications. It can be linked directly
or at runtime using the \texttt{LD\_PRELOAD} mechanism. \doubletake{}
is thus convenient to use: there is no need to change or recompile
applications, to use a specialized hardware platform, run inside a
virtual machine, or modify the OS.
%\doubletake{} is very convenient. 
%, or can be injected into unmodified binaries by setting the \texttt{LD\_PRELOAD} environment variable on Linux.

At startup, \doubletake{} begins the first epoch. This epoch continues
until the program issues an \emph{irrevocable} system call (see
Section~\ref{sec:syscalls} for details). Before an irrevocable system
call, \doubletake{} checks program state for evidence of errors. The
details are presented in Section~\ref{sec:applications}.

If no errors are found, \doubletake{} ends the current epoch, issues
the irrevocable system call, and begins a new epoch. If it finds
evidence of an error, \doubletake{} enters re-execution
mode. \doubletake{} will then re-execute with instrumentation
activated and report the lines of code responsible for the
error(s).

The remainder of this section describes the implementation
of \doubletake{}'s core functionality.

\subsection{Startup and Shutdown}

% Memory initialization. Get real functions. Or does some other things. 
At program startup, \doubletake{} performs initialization and starts
the first epoch. \doubletake{} needs to get in early to interpose on
system calls and install its own heap implementation. It
accomplishes this by marking its own initialization function with the
constructor attribute. Since \doubletake{} must wrap library functions
that eventually invoke with system calls, as described in
Section~\ref{sec:syscalls}, it collects the addresses of all
intercepted functions during this initialization phase. \doubletake{}
acquires memory from the OS to hold its heap, collects the
names and ranges of all globals by analyzing \texttt{/proc/self/maps},
installs signal handler for segmentation violations, and prepares the
data structure for recording and handling system calls.

For technical reasons, \doubletake{} must postpone the checkpointing
of program state (and thus the beginning of the first epoch) until
just before execution enters the application enters
its \texttt{main} function. This delay is necessary to let key low-level
startup tasks complete. For example, C++ performs its initialization
for the standard stream after the execution of constructor functions
(including, in this case, \doubletake{} itself). Because \doubletake{}
relies on streams to report any errors it detects, by definition it
cannot start the first epoch before that point. To make this all possible,
we interpose on the \texttt{libc\_start\_main} function, and pass a
custom \texttt{main} function implemented by \doubletake{} that
performs a snapshot just before entering the application's real
\texttt{main} routine.

\doubletake{} treats program termination as the end of the final epoch. As with any other epoch, if it finds evidence of program
errors, \doubletake{} re-executes the program to pinpoint the exact
causes of errors. This logic is embedded in a finalizer marked with the
deconstructor attribute that \doubletake{} installs.

\newpage

\subsection{Epoch Start}
\label{sec:implementation/start}

At the beginning of each epoch, \doubletake{} takes a snapshot of
program state. \doubletake{} saves all writable memory (stack, heap,
and globals) from the main program and any linked libraries, and saves
the register state of each thread with the \texttt{getcontext}
function. To reduce the cost of snapshots, \doubletake{} does not
checkpoint any read-only memory. To identify all writable mapped
memory, \doubletake{} processes the \texttt{/proc/self/map} file,
which on Linux identifies every mapped memory region and its
attributes (other operating systems implement similar
functionality). \doubletake{} also records the file positions of all
open files, which lets programs issue \texttt{read} and \texttt{write}
system calls without ending the current epoch. \doubletake{} uses the
combination of saved memory state, file positions and registers to
rollback execution if it finds evidence of an error.

% About the start of a program

% About the initialization function. 
%%%%%%%%%%%%%%%%%%%%%%%%%%%

\subsection{Normal Execution}
\label{sec:implementation/normalexecution}

Once a snapshot has been written, \doubletake{} lets the program
execute normally but interposes on heap allocations/deallocations and
system calls in order to set tripwires and support re-execution.

%%%%%%%%%%%%%%

\subsubsection*{System Calls}
\label{sec:syscalls}
\begin{table}[t]
	\centering
	\small
	\renewcommand{\arraystretch}{1.5}
	\begin{tabular}{l|p{6cm}}
		\textbf{Category} & \textbf{Functions} \\
		\hline
		
		\emph{\textbf{Repeatable}}		& \texttt{getpid}, \texttt{sleep}, \texttt{pause}\\
		
		\emph{\textbf{Recordable}}		& \texttt{mmap}, \texttt{gettimeofday}, \texttt{time}, 
						  \texttt{clone} , \texttt{open}\\
		
		\emph{\textbf{Revocable}}		& \texttt{write}, \texttt{read} \\
		
		\emph{\textbf{Deferrable}}		& \texttt{close}, \texttt{munmap} \\
		
		\emph{\textbf{Irrevocable}}		& \texttt{fork}, \texttt{exec}, \texttt{exit}, \texttt{lseek}, \texttt{pipe}, \texttt{flock}, \texttt{socket related system calls}\\
	\end{tabular}
	\caption{System calls handled by \doubletake{}. All unlisted system calls are conservatively treated as irrevocable, and will end the current epoch. Section~\ref{sec:syscalls} describes how \doubletake{} handles calls in each category.\label{table:syscalls}}
\end{table}

\doubletake{} ends each epoch when the program attempts to issue an
irrevocable system call. However, most system calls can safely be
re-executed or undone to enable re-execution.

\doubletake{} divides system calls into five categories, shown in
Table~\ref{table:syscalls}. System calls could be intercepted
using \texttt{ptrace}, but this would add unacceptable overhead during
normal execution. Instead, \doubletake{} interposes on all library
functions that may issue system calls.

\begin{itemize}

%%%%%%%

\item
\textbf{Repeatable system calls} do not modify system state, and return
the same result during normal execution and re-execution. No special
handling is required for these calls.
%\todo{If you don't guarantee perfect replay, then you must argue that perfect replay isn't necessary and that at the very least the attempted replay does not affect application behavior.}

%%%%%%%

\item
\textbf{Recordable system calls} may return different results if they
are re-executed. \doubletake{} records the result of these system
calls during normal execution, and returns the saved result during
re-execution. Some recordable system calls, such as \texttt{mmap},
change the state of underlying OS.

\item
\textbf{Revocable system calls} modify system state, but \doubletake{}
can save the original state beforehand and restore it prior to
re-execution. Most file I/O operations fall into this category. For
example, although \texttt{write} modifies file contents, \doubletake{}
can write the same content during re-execution. The \texttt{write} function
also changes the current file position, but the file position can be
restored to the saved one using \texttt{lseek} prior to
re-execution.

At the beginning of each epoch, \doubletake{} saves all file
descriptors of opened files in a hash table. Maintaining this hash
table helps to identify whether a \texttt{read} and \texttt{write}
call is operating on sockets or not, because socket communications
must be treated as irrevocable system calls. In
addition, \doubletake{} must save stream contents returned
by \texttt{fread} in order to support re-execution.

\item	
\textbf{Deferrable system calls} will irrevocably change program state,
but can safely be delayed until the end of the current
epoch. \doubletake{} delays all calls to \texttt{munmap}
and \texttt{close}, and executes these system calls before
starting a new epoch when there is no need to re-execute the program.
	
\item
\textbf{Irrevocable system calls} change internally-visible program
state, and cannot be rolled back and re-executed. \doubletake{}
ends the current epoch before these system calls.

\end{itemize}

\doubletake{} reduces the number of irrevocable system calls by
observing their arguments; in some cases, they are not necessarily
irrevocable. For example, when \texttt{fcntl} invoked
with \texttt{F\_GET}, \doubletake{} treats it as a {\it repeatable
system call} since it is simply a read of file system state. However,
it treats this call as irrevocable if invoked with \texttt{F\_SET},
since the call then actually updates the file system.
%%%%%%%%%%%%%%

%\input{multithreading}

\subsubsection*{Memory Management}
\label{sec:heapallocator}

%% Save this multithreading support for the future. 
%\doubletake{}'s heap is completely deterministic for a program without races, so no logging is required to ensure that allocations do not change during re-execution. To achieve this target, all allocations from the same thread is met at a specific subheap, which obtains every superblock (large chunks of memory) by acquiring a lock. Superblock allocations are deterministic in replay phase, which is enforced by the deterministically acquiring and releasing of a lock. When an object is freed, this object is returned to the current thread issuing this free operation. Because of deterministic superblock allocations and memory free operations (enforced by the program order), we guarantee that all memory allocations are deterministic inside the same subheap. 

As described in Section~\ref{sec:heap}, \doubletake{} intercepts
memory allocations and deallocations to implant \emph{tripwires},
identify heap corruption, and facilitate re-execution. \doubletake{}
replaces the default heap with a fixed-size BiBOP-style allocator with
per-thread subheaps and power-of-two size classes. We built this heap using the {\scshape Heap
Layers} framework~\cite{heaplayers}.

\doubletake{} implants tripwires differently for different
analyses. To detect heap-based buffer overflows, \doubletake{} places
canaries along with each heap object. In order to find use-after-free
errors, \doubletake{} postpones the reuse of freed objects by putting
them into a quarantine list and filling them with canaries. For memory
leak detection, there is no need to implant tripwires, because the
evidence of a leak can be found without them.
 
To identify heap corruption, \doubletake{} maintains a bitmap that
records the locations of all heap canaries. The bitmap records every
word of heap memory that contains a canary, which will be checked at
the end of each epoch. If any of these words are
modified, \doubletake{} notifies the detection tool.

%To correctly identify buffer overflows of objects with size less than power of 2, \doubletake{} keeps size and status information of each heap object at their own object headers. 
%During allocation, objects are marked as allocated and their \emph{requested size} are saved, which may be less than the power-of-two size class. At deallocation, objects are marked as freed and placed into the quarantine list if the detection of use-after-free is enabled. For non-aligned objects, \doubletake{} also checks buffer overflows for non-aligned words, which can even report one-byte buffer overflows because of exact size. 

%\doubletake{} also Re-execution is only started if the detection tool finds that canaries between allocated objects have been overwritten.

%During replay, \doubletake{}'s heap allocator checks to see if the object being allocated or freed contains the address where an error was detected. If so, \doubletake{} calls the \texttt{backtrace()} function to obtain a call stack for the allocation and deallocation sites. 

To speed re-execution, \doubletake{} uses its heap
allocator to satisfy memory requests from the application and
corresponding libraries, and maintains a separate heap for internal
use only. For example, the memory that \doubletake{} uses to record
system calls results is allocated from its internal heap and there is
no need to replay these allocations during re-execution. Any
additional memory allocations during the replay phase are also
satisfied from its internal heap.

\subsection{Epoch End}

Each epoch ends when the program issues an irrevocable system call. 
%All other threads are notified with a signal SIGUSR2. Once all threads have stopped, 
At the end of each epoch, \doubletake{} checks program state for
errors. These analysis-specific error checks are described in
Section~\ref{sec:applications}. If an error is found, \doubletake{}
rolls back execution to the immediately-preceding epoch, and switches
to re-execution mode. If no error is found, \doubletake{} issues any
deferred system calls, clears the logs for all recorded system calls,
and begins the next epoch.

%%%%%%%%%%%%%%%%%%%%%%%%%%%

\subsection{Rollback}

If an error is found, \doubletake{} rolls back program state prior to
beginning re-execution. This rollback must be handled with care.

For example, restoring the saved stack may corrupt the current stack
if the size of the saved stack is larger than that of the current
stack. \doubletake{} thus switches to a temporary stack during its
rollback phase. When performing rollback, the saved state of all
writable memory is copied back, which also recovers the status of its heap. \doubletake{} also recovers the file positions of opened
files so that all read/write calls can be issued normally during
re-execution.

\doubletake{} then sets hardware watchpoints on all corrupted
addresses in order to report the root causes of buffer overflows or
dangling pointers. Since debug registers are not directly accessible
in user mode, \doubletake{} utilizes the \texttt{perf\_event\_open}
call to load watched addresses into the debug registers. \doubletake{}
also sets a \texttt{SIGTRAP} handler for watchpoints so that it will
get notified when these addresses are overwritten (e.g., during buffer
overflows or uses of freed objects).

Once all watchpoints have been placed, \doubletake{} uses
the \texttt{setcontext} call to restore register state and begin
re-execution.

\subsection{Re-Execution}
\label{sec:implementation/re-execution}

During re-execution, \doubletake{} replays the saved results of
recordable system calls from the log collected during normal
execution, while avoiding invoking actual system calls; that is, their
execution is simulated. All deferred system calls are converted to
no-ops while the program is re-executing. \doubletake{} issues other
types of system calls normally.

\doubletake{}'s heap design and its rollback has recovered the memory
state to the snapshotted state. To repeat the replayable memory
uses, \doubletake{} simply repeats memory allocations and
deallocations from applications and libraries according to the program
order. The additional memory uses happened in the replay phase, such
as bookkeeping the call stack of memory uses, will be satisfied
from \doubletake{}'s internal heap and will not affect the memory uses
of applications.

During replay, \doubletake{} enables tracking of precise information
in the memory allocator: all allocations and deallocations record
their calling context so these can be reported later, if needed. Note
that recording call sites during ordinary execution would be prohibitively
expensive, imposing 20--30\% overhead; \doubletake{}'s strategy removes
this overhead from normal execution.

Finally, \doubletake{} handles traps caused by accesses to
watchpoints. Inside the trap handler, \doubletake{} first determines
which watchpoint caused the current trap if there are multiple
watchpoints. It also filters out any accesses from \doubletake{}
itself. \doubletake{} prints the callsite stack of the instruction
responsible for a buffer overflow or use-after-free errors and their
memory allocation (or deallocation) sites. For memory
leaks, \doubletake{} reports the allocation callsite of the leaked
object.

%Normally, the debug status register has to be accessed in order to obtain this information. However, using \texttt{ptrace} is inconvenient inside a signal handler because of involving in another process. \doubletake{} always maintains an updated value for every watchpoint. Thus, it can precisely determine which watchpoint is triggered by checking the changes of those watchpoints. 

%\section{\doubletake{} for Multithreading Programs}
%\input{multithreading}

%\section{Optimization}
%\input{optimization}

\section{Evaluation}
\begin{figure*}[ht!]
	\begin{center}
		\includegraphics[width=6.5in]{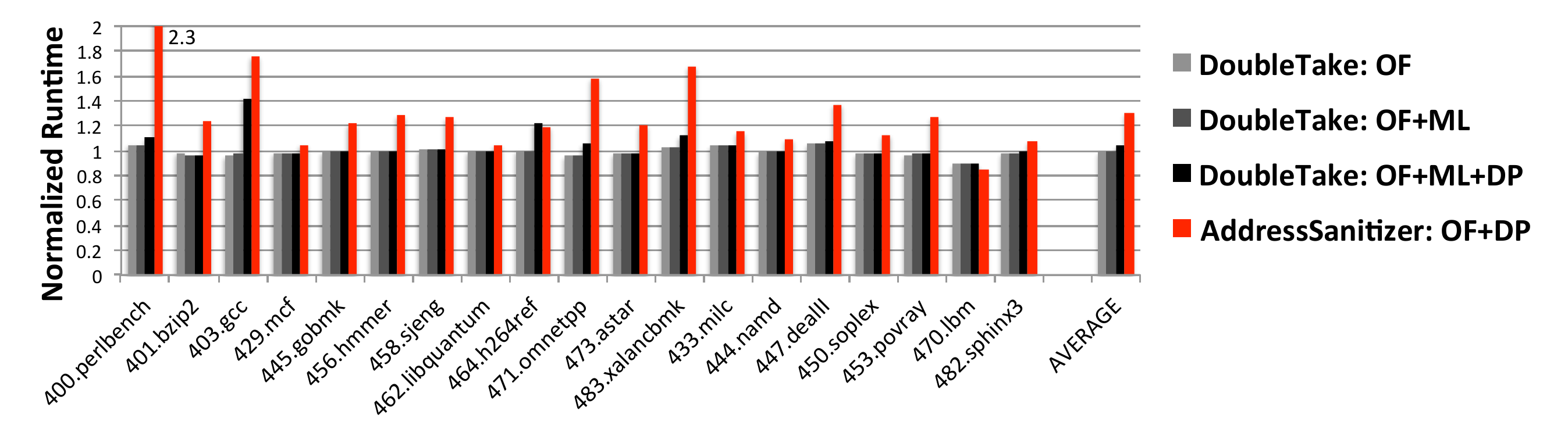}
	\end{center}
	\caption{Runtime overhead of \doubletake{} (OF = Buffer Overflow Detection, ML = Memory Leak Detection, DP = Dangling Pointers Detection) and AddressSanitizer, normalized to each benchmark's original execution time. With all detections enabled, \doubletake{} only introduces 4\% performance overhead  on average.
%Overhead for Valgrind is reported in Table~\ref{table:valgrind} because the results do not fit on this graph.
\label{fig:perf}}
\end{figure*}

\label{sec:evaluation}
% Explain why we don't evaluate on parallel programs

We evaluate \doubletake{} to demonstrate its efficiency, in terms of
execution time, memory overhead, and effectiveness at detecting
errors. All experiments are performed on a quiescent Intel Core 2
dual-processor system with 16GB of RAM, running on Linux
3.13.0-53-generic with \texttt{glibc-2.19}. Each processor is a 4-core
64-bit Intel Xeon, operating at 2.33GHz with a 4MB shared L2 cache and
a 32KB per-core L1 cache. All programs are built as 64-bit executables
using LLVM 3.2 with the clang front-end and \texttt{-O2}
optimizations. All evaluations on SPEC CPU2006 are exercised with the
``ref'' (reference) input set.

% -194.17.1.el5

\subsection{Runtime Overhead}
\label{sec:evaluation/runtime}

We evaluate \doubletake{}'s runtime and memory overhead across all of
the C and C++ SPEC CPU2006 benchmarks, 19 in total. We
compare \doubletake{} with the previous state-of-the-art tool,
Google's AddressSanitizer~\cite{AddressSanitizer}.  As mentioned
earlier, AddressSanitizer can detect buffer overflows and
use-after-free errors, but it only detects memory leaks at the end of
execution. By contrast, \doubletake{} detects all of these errors at
the end of every epoch.

% We therefore configure AddressSanitizer to detect only
%buffer overflows and use-after-free errors on heap
%objects. \footnote{\small We invoke AddressSanitizer with
%the following configuration flags:
%\texttt{-O2 -fsanitize=address -g -mllvm -asan-stack=0 -mllvm -asan-opt=1 -mllvm -asan-instrument-reads=0
%-mllvm -asan-instrument-writes=1 -mllvm -asan-mapping-scale=3 -mllvm -asan-mapping-offset-log=-1 -mllvm -asan-use-after-return=1 -mllvm -asan-globals=0 -m64}}. 

In our evaluation, \doubletake{} discovered several memory leaks,
which trigger rollback and error identification. To isolate normal
execution overhead, we disable \doubletake{}'s rollback in our
evaluation. That is, our runs with \doubletake{} incur all of the
overhead of ordinary tracking (including implanting of tripwires and
examining state) but do not measure the time to rollback and locate
errors; in general, this cost is low and in any event does not affect
bug-free execution, which is the common case. For each benchmark, we
report the average runtime of three runs.

%Because Valgrind's overhead is so high (between 4.5$\times$ and 42.8$\times$), we exclude it from the graph (see Table~\ref{table:valgrind} for details). 

Figure~\ref{fig:perf} presents execution time overhead results
for \doubletake{} and AddressSanitizer.  On average, \doubletake{}
imposes only $4\%$ overhead \emph{with all three error detectors
enabled}. When use-after-free detection (DP) is
disabled, \doubletake{} exhibits no observable
overhead. AddressSanitizer has an average runtime overhead over
$30\%$; recall that AddressSanitizer only performs leak detection at
the end of program execution, while \doubletake{} performs it every
epoch.

% difference across all different tools

For 17 out of 19 benchmarks, \doubletake{} outperforms
AddressSanitizer.  For 14 benchmarks, \doubletake{}'s runtime overhead
with all detectors enabled is under 3\%. Unsurprisingly,
both \doubletake{} and AddressSanitizer substantially outperform
Valgrind on all benchmarks.

Four of the benchmarks have higher than average
overhead for \doubletake{} and
AddressSanitizer (\texttt{400.perlbench}, \texttt{403.gcc}, \texttt{464.h264ref},
and \texttt{483.xalancbmk}). Both \doubletake{}
and AddressSanitizer substantially increase these applications' memory footprints
(see Table~\ref{tbl:memoryoverhead}). We attribute their
increased execution time to this increased memory footprint and its
corresponding increased cache and TLB pressure.

%For \texttt{gcc}, \doubletake{} introduces 40\% performance overhead, with use-after-free detection enabled, because of delaying memory uses that greatly increases the footprint of programs. 

% Difference across all different tools
\doubletake{}'s use-after-free detection adds roughly $4\%$ runtime
overhead, but only \texttt{gcc} and \texttt{h264ref} run with more
than $20\%$ overhead. As described in
Section~\ref{sec:applications/useafterfree}, all freed objects are
filled with canaries (up to 128 bytes). \doubletake{} spends a
substantial amount of time filling freed memory with canaries for
applications with a large number of \texttt{malloc} and \texttt{free}
calls. Thus, \doubletake{} runs much slower for the
application \texttt{gcc} when the detection of use-after-free errors is
enabled. \texttt{h264ref} adds significant overhead on \doubletake{}
because of its large number of epochs.

Table~\ref{table:character} presents detailed benchmark
characteristics. The ``Processes'' column shows the number of
different process invocations (by calling \texttt{fork}). The number
of epochs is significantly lower than the number of actual system
calls, demonstrating \doubletake{}'s effectiveness at reducing epochs
via its lightweight system call handling. The benchmarks with the
highest overhead share the following characteristics: they consist of
a substantial number of epochs (e.g., \texttt{perlbench}
and \texttt{h264ref}) or are unusually \texttt{malloc}-intensive
(e.g., \texttt{gcc}, \texttt{omnetpp}, and \texttt{xalancbmk}).

\textbf{Runtime Overhead Summary:}
For nearly all of the benchmarks we examine, \doubletake{}
substantially outperforms the state of the art. For most benchmarks, \doubletake{}'s runtime
overhead is under 3\%.

%%%%%%%%%%%%%%%%%%%%%%%%%%%%%%%

\begin{figure*}[ht!]
\begin{center}
\includegraphics[width=6.5in]{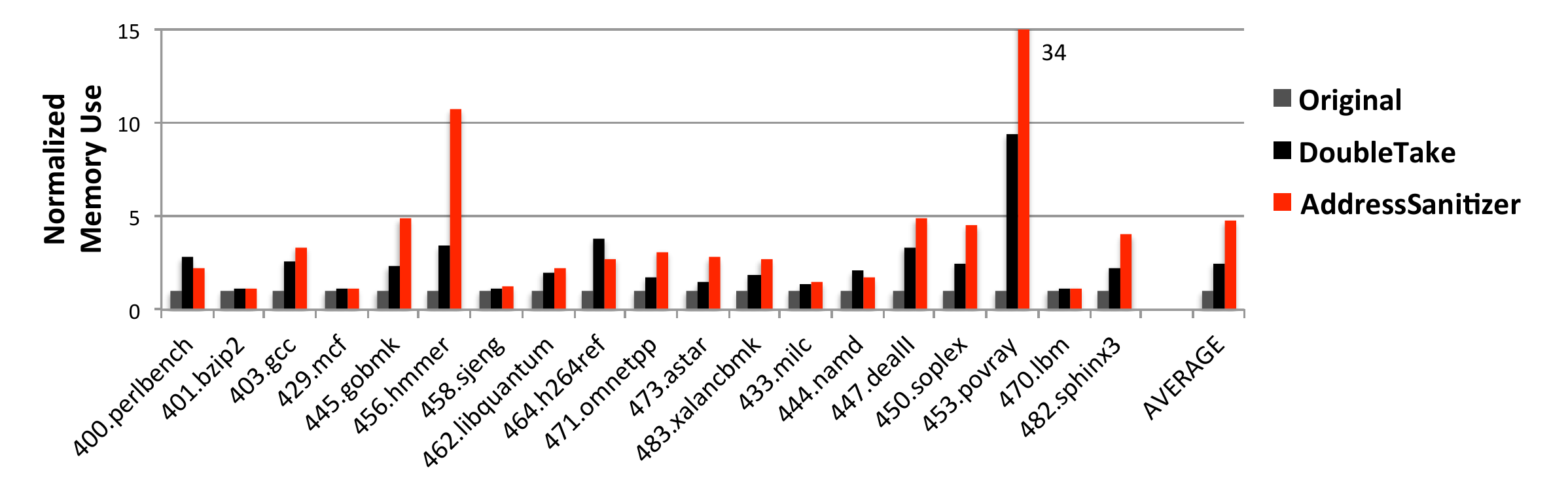}
\end{center}
\caption{
Memory overhead of \doubletake{} and AddressSanitizer.
\label{fig:memory}}
\end{figure*}

\subsection{Memory Overhead}
\label{sec:evaluation/memory}
We measure program memory usage by recording the peak \emph{physical}
memory usage. Virtual memory consumption is generally not relevant for
64-bit platforms, which have enormous virtual address
ranges. \doubletake{}'s pre-allocated heap and internal heap consume
8GB of virtual memory space. We compute peak physical memory usage by
periodically collecting process-level information (on Linux, this is available in the 
\texttt{/proc/self/smaps} pseudo-file), and summing the proportional
set sizes of memory
segments. % ~\cite{memuses}.

Figure~\ref{fig:memory} presents memory overhead for \doubletake{} and
AddressSanitizer (Table~\ref{tbl:memoryoverhead} has full details).
On average, both across the benchmark suite
and when broken down by footprint (large ($>$ 100MB) and small ($<$
100MB)), \doubletake{} imposes considerably lower memory overhead than
AddressSanitizer. \doubletake{} imposes lower
memory overhead than AddressSanitizer on all but three
benchmarks: \texttt{perlbench}, \texttt{h264},
and \texttt{namd}.

\punt{
%However, the numbers look different when we separate out small and large-footprint applications. 
From Table~\ref{tbl:memoryoverhead}, we can see that \doubletake{}
imposes lower memory overhead for large-footprint
applications, with one exception
(\texttt{447.dealII}). Both \doubletake{} and AddressSanitizer have
relatively high memory overhead for small footprint applications
like\texttt{453.povray} and \texttt{464.h264ref}, since these
benchmarks originally only use just 3MB and 24MB, respectively.  For
all other benchmarks, both \doubletake{} and AddressSanitizer
introduce less than $5\times$ memory overhead.
}

We drill down to explore the application and analysis
characteristics that contribute to \doubletake{}'s memory overhead:

\begin{itemize}
\item \textbf{Number of epochs:} 
Much of \doubletake{}'s memory overhead comes from the snapshot of
writable memory taken at the beginning of each epoch. However, the first
snapshot is often small because the heap is almost empty
before the \texttt{main} routine. For example, the
benchmarks \texttt{bzip2}, \texttt{mcf}, \texttt{sjeng}, \texttt{milc},
and \texttt{lbm} run in a single epoch, and accordingly exhibit very low
memory overhead. 

\item \textbf{System call logs:}
System call logs introduce additional memory overhead
that depends on the number of recorded system calls.

\item \textbf{Analysis-specific overhead:}
Other sources of memory overhead are analysis-specific. Buffer
overflow detection adds canaries between heap objects, which can
increase memory usage for programs with many small allocations. For
this analysis, \doubletake{} also maintains a bit map that marks the
placement of canaries: for each eight-bytes word of the
heap, \doubletake{} adds a bit to mark whether this word has the
canaries or not. Finally, use-after-free detection adds constant-size
memory overhead by delaying memory reuse. Note that any similar dynamic analyses
must impose similar overheads.
\end{itemize}

\begin{table}[!t]
\small
\centering
\begin{tabular}{lrrr}
\multicolumn{4}{c}{\textbf{\small Memory Usage (MB)}} \\
\hline
\textbf{ \small Benchmark} & \textbf{\small Original} &  \textbf{\small Address\-} & \textbf{\small \scshape{Double}} \\
 &  & \textbf{\small Sanitizer} & \textbf{\small \scshape{Take}} \\
\hline
\multicolumn{4}{c}{\textit{large footprint ($>$ 100MB)}} \\
400.perlbench & 656 &	1481 & 1834 \\
401.bzip2	& 870 &	1020 &	989 \\
403.gcc	& 683 &	2293 &	1791      \\
429.mcf	& 1716 &	1951  &	2000 \\
458.sjeng & 179 & 220 &	212  \\
471.omnetpp	& 172 &	538  &	299  \\
473.astar	& 333 &	923  &	479 \\
483.xalancbmk   & 428 & 1149 &	801  \\
433.milc	& 695 &	1008 &	917  \\
447.dealII	& 514 &	2496 &	1724  \\
450.soplex	& 441 &	1991 &	1104 \\ 
470.lbm & 418 &	496 &	477 \\
\textbf{geometric mean} & & \emph{+117\%} & \emph{+72\%} \\ \hline 
\multicolumn{4}{c}{\textit{small footprint ($<$ 100MB)}} \\
445.gobmk &	28 &	137 &	66 \\
456.hmmer &	24 &	256 &	82 \\
462.libquantum	& 66 &	144 &	132 \\
464.h264ref	& 65 &	179 &	247 \\
444.namd	& 46 &	79 &	96 \\
453.povray	& 3 &	133 &	37 \\
482.sphinx3     &	45 &	181 & 103 \\
\textbf{geometric mean} & & \emph{+395\%} & \emph{+208\%} \\
\hline
% \textbf{Total} & \textbf{7386} & \textbf{16678} & \textbf{13391} \\
\textbf{overall geometric mean} &         & \emph{+194\%}  & \emph{+114\%} \\
\end{tabular}
\caption{AddressSanitizer and \doubletake{} memory usage, in megabytes. The top section lists memory overhead for large-footprint applications (over 100 MB), while the bottom section presents overhead for small-footprint applications. \doubletake{}'s memory overhead is generally less than AddressSanitizer's.} \label{tbl:memoryoverhead}
\end{table}

\textbf{Memory Overhead Summary:}
On average, \doubletake{} imposes lower memory overhead than
AddressSanitizer. For large footprint applications, it increases memory consumption
for applications by 72\% on average.

%%%%%%%%%%%%%%%%%%%%%%%%%%%%%%%

\subsection{Effectiveness}
\label{sec:effect}

We evaluate the effectiveness of \doubletake{} on a range of
applications, including synthetic test cases, standard benchmarks, and
real-world applications.
 
\textbf{Synthetic test cases and benchmarks:} 
%\vspace{0.1in}
We first evaluate \doubletake{} on 3 synthetic test cases, 26 test
cases from the NIST SAMATE Reference Dataset Project. This corpus
includes 14 cases with buffer overflows and 12 cases without
overflows~\cite{microbenchmarks}. We also evaluate \doubletake{} on 19
C/C++ benchmarks from the SPEC CPU2006 benchmark suite.

For heap overflows, \doubletake{} detects all known overflows in one
synthetic test case and 14 test cases of SAMATE suite. For the 12
cases without overflows in SAMATE suite, \doubletake{} has no false
positives. For the SPEC CPU2006 benchmarks, \doubletake{} did not find
any heap buffer overflows and use-after-frees, which is the same
result found with AddressSanitizer. However, \doubletake{} detected a
significant number of memory leaks in \texttt{perlbench}
and \texttt{gcc} of SPEC CPU2006, which we verified using
Valgrind's Memcheck tool.

%\doubletake{} detected a lot of buffer overflows in SPEC CPU2006 benchmark suite, which cannot be detected by AddressSanitizer. Those buffer flows are listed in the Table~\ref{}. 
% SAMATE: 
% Doubletake: 13
% AddressSanitizer: 
% heap_overflow_cplx_good_1846-pthread: (crash for both) 
% heap_overflow_location_good_1848-pthread: false positives (should be bad now)
 
\begin{table}
  \small
  \centering
  \begin{tabular}{l | l | l | l}
   \textbf{Application} & \textbf{Description}  & \textbf{LOC} & \textbf{Error Type} 
   \\ \hline
   \small{bc} & \small{basic calculator} & \small{12K} & \small{Known Overflow} \\
   \small{gzip} & \small{compress or expand files} & \small{5K} & \small{Converted Overflow} \\
   \small{libHX} & \small{common library} & \small{7K} & \small{Known Overflow} \\
   \small{polymorph} & \small{filename converter} & \small{0.4K} & \small{Converted Overflow} \\
  \small{vim-6.3} & \small{text editor} & \small{282K} & \small{Known Overflow} \\
  \small{gcc-4.7} & \small{GNU Compiler Collection} & \small{5784K} & \small{Unknown leaks} \\
  \small{vim-6.3} & \small{text editor} & \small{282K} & \small{Unknown leak} \\
  \small{ls} & \small{directory listing} & \small{3.5K} & \small{Implanted UAF} \\ 
  \small{wc} & \small{word count} & \small{0.6K} & \small{Implanted UAF} \\
  \small{vim-7.4} & \small{text editor} & \small{332K} & \small{Implanted UAF} \\
   \hline
   
   \end{tabular}
  \caption{
    Error detection: \doubletake{} detects both known (injected and non-injected) and previously unknown errors on the above applications (any reported errors are real, as \doubletake{} has a zero false positive rate).\label{fig:listoferrors}}
\end{table}

\textbf{Real applications:} 
%\vspace{0.1in}
We also ran \doubletake{} with a variety of applications with known
errors or implanted errors, listed in Table~\ref{fig:listoferrors}.
To verify the effectiveness of \doubletake{}'s buffer overflow
detection, we collected applications from evaluations of prior buffer
overflow detection tools, Bugzilla, and
bugbench~\cite{vimoverflow, bzip2overflow,bugbench,overflow:Cruiser},
including \texttt{bc}, \texttt{gcc-4.4.7}, \texttt{gzip}, \texttt{libHX}, \texttt{polymorph},
and \texttt{vim-6.3}.

In every case, \doubletake{} detected all known or converted
errors. Converted errors are existing global or array overflows
that \doubletake{} currently cannnot detect; we converted these to
heap overflows to verify its effectiveness. \doubletake{} also
identified memory leaks in \texttt{gcc-4.4.7} and \texttt{vim-6.3},
which we confirmed with Valgrind. To evaluate the detection of
use-after-free errors, we manually injected errors on real
applications, such as \texttt{vim-7.3}, \texttt{ls}
and \texttt{wc}. \doubletake{} identified all of these memory
errors.

Note that the errors observed in these applications are triggered only
by specific inputs. In the common case, these applications perform as
expected. This is exactly the case for which \doubletake{} is ideal,
since its low overhead is designed to make it feasible to use it in
deployed settings.

%\begin{figure*}[htbp]
\begin{figure*}[t!]
%\centering
\subfigure[\doubletake{} Report]{%
   \label{fig:doubletakereport}
    \includegraphics[width=3.4in]{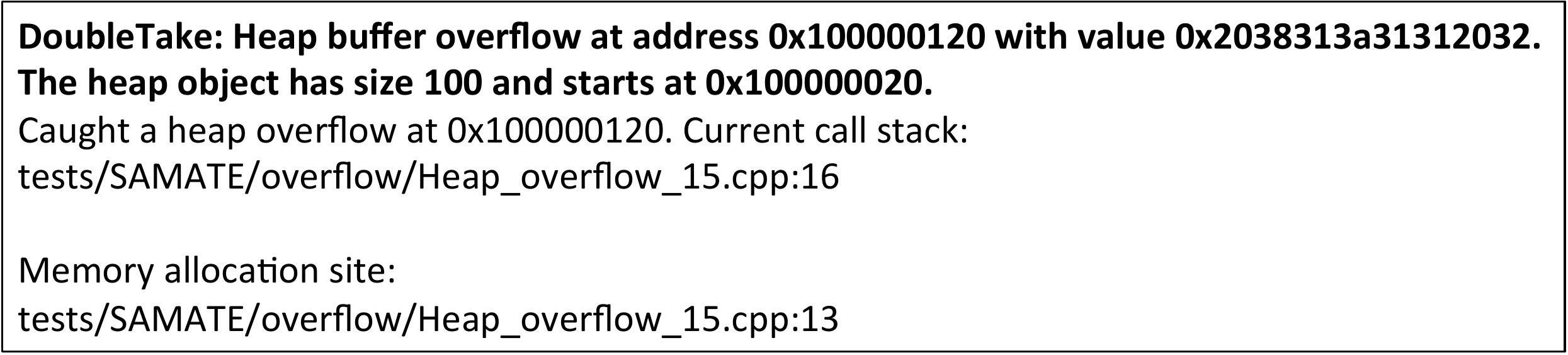}

}%
%\vspace{15pt}
\subfigure[AddressSanitizer Report]{%
   \label{fig:asanreport}
   \includegraphics[width=3.4in]{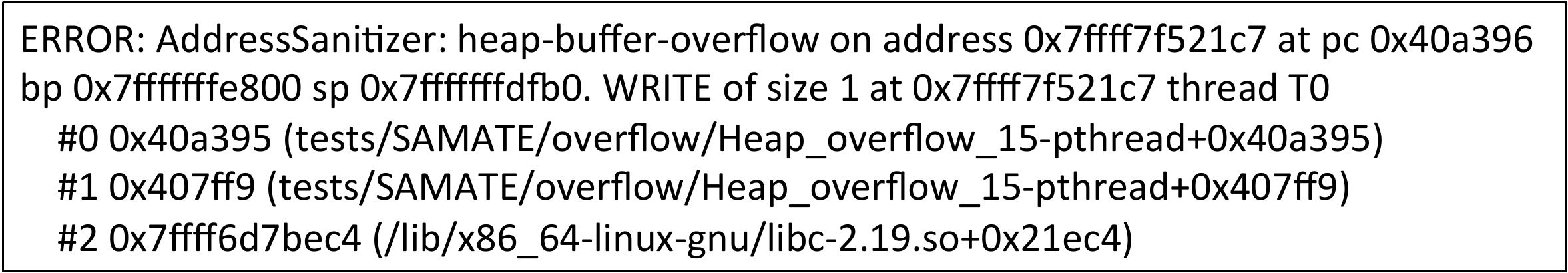}
}
\caption{
Example reports of \doubletake{} and AddressSanitizer for buffer overflow identification.
\label{fig:reports}}
\end{figure*}

\textbf{Detailed reporting:}
\doubletake{} reports precise information aimed at helping programmers
identify the exact causes of different memory errors, as shown in
Figure~\ref{fig:doubletakereport}. For buffer overflows, \doubletake{}
reports the call sites and line numbers of the overflow and the
original memory allocation. For memory leaks, \doubletake{} reports
the last call site of its memory allocation. For use-after-frees
error, \doubletake{} reports both allocation and deallocation call
sites, and the instruction(s) that wrote to the object after it was
freed. In general, \doubletake{} provides more detailed information
than AddressSanitizer, as seen in Figure~\ref{fig:asanreport}.

In addition, \doubletake{} can identify more errors than
AddressSanitizer. \doubletake{} can track up to four buffer overflows
or use-after-free errors during the same epoch because its isolation
is based on the use of hardware debugging registers. AddressSanitizer
always stops at the detection of the first such error.

\textbf{Effectiveness Summary:}
Across the applications we examine, \doubletake{} detects all known or
injected errors with no false positives. \doubletake{} is as effective
at finding errors as AddressSanitizer, but with much lower performance
and memory overhead. It also provides more detailed reports for these
errors.

\begin{table}[!t]
\small
\centering
\begin{tabular}{l|r| r| r|r }
\textbf{Benchmark} & \textbf{Processes} & \textbf{Epochs} & \textbf{Syscalls} & \textbf{\# Mallocs} \\
\hline
400.perlbench & 3 & 43 & 60068 & 360605640 \\
401.bzip2 & 6 & 6 & 968 & 168 \\
403.gcc & 9 & 9 & 155505 & 28458514 \\
429.mcf & 1 & 1 & 24443 & 5 \\
445.gobmk & 5 & 5 & 2248 & 658034 \\
456.hmmer & 2 & 2 & 46 & 2474268 \\
458.sjeng & 1 & 1 & 23 & 5 \\
462.libquantum & 1 & 1 & 11 & 179 \\
464.h264ref & 3 & 825 & 2592 & 146827 \\
471.omnetpp & 1 & 1 & 19 & 267168472 \\
473.astar & 2 & 2 & 102 & 4799955 \\
483.xalancbmk & 1 & 1 & 123706 & 135155557 \\
433.milc & 1 & 1 & 12 & 6517 \\
444.namd & 1 & 1 & 470 & 1324 \\
447.dealII & 1 & 1 & 8131 & 151332314 \\
450.soplex & 2 & 2 & 37900 & 310619 \\
453.povray & 1 & 1 & 25721 & 2461141 \\	
\end{tabular}
\caption{Benchmark characteristics. \label{table:character}}
\end{table}

% {\bf Note:} Although AddressSanitizer claims that they cannot detect buffer overflows or use-after-frees with less-than-one-word access~\cite{AddressSanitizer}, we do not find this problem during our evaluation. Both \doubletake{} and AddressSanitizer are both effective in detecting these two types of errors, with the difference only in the performance and reporting information. 

%\doubletake{}'s performance overhead for these applications was either low or unnoticeable.
%Several of these applications are interactive (\texttt{bc}
%and \texttt{vim}), and the use of \doubletake{} had no perceptible
%impact.

%\textbf{Need to quantify overhead.}

\section{Discussion}
\label{sec:discuss}

% \subsubsection*{Analyses}
The analyses we have built using \doubletake{} (heap buffer overflows,
use-after-free errors, and memory leaks) have no false positives, but
they can have false negatives. Our heap buffer overflow detector
cannot identify all non-contiguous buffer overflows, a limitation of
as all canary-based detectors. If an overflow touches memory only in
adjacent objects and skips over canaries, \doubletake{}'s end-of-epoch
scan will not reveal any evidence of the overflow. Both the buffer
overflow and use-after-free detectors can detect errors only on
writes. To reduce overhead, the use-after-free detector only places
canaries in the first 128 bytes of freed objects. If a write to freed
memory goes beyond this threshold, our detector will not find it. The
memory leak detector will not produce false positives, but non-pointer
values that look like pointers to leaked objects can lead to false
negatives. Finally, if a leaked object was not allocated in the
current epoch, \doubletake{}'s re-execution will not be able to find
the object's allocation site (a limitation shared by
AddressSanitizer). In practice, \doubletake{}'s epochs are long enough
to collect allocation site information for all leaks detected during
our evaluation.

%In this section, we discuss limitations of \doubletake{} and the
%detection tools we have implemented. %, as well as plans for future work.

While evidence-based dynamic analyses can run with very low overhead,
they cannot detect errors if there is no evidence, or it is not
practical to force evidence of their existence. Evidence-based
analysis also depends on errors being generally monotonic: once an
error has occurred, its evidence needs to persist until the end of
the epoch in order to ensure detection.

Finally, the current prototype of \doubletake{} is limited
to executing single-threaded code. However, we believe this is primarily an
engineering question. Evidence-based analysis does not depend on fully
deterministic replay. Consider the case
of a memory error arising due to a race. During replay, the same sequence of
writes may not recur, and thus the hardware watchpoints in those
addresses might not be triggered. In this scenario, \doubletake{} can
simply continue execution having successfully masked the error, or
repeatedly re-execute the epoch in an effort to expose the data
race. Because of this flexibility, \doubletake{} will not need to
track details about the ordering of memory accesses, which is what
makes deterministic record-and-replay systems expensive. Supporting
multithreaded programs will only require interception of
synchronization operations to allow \doubletake{} to pause threads at
epoch boundaries and to track the order of synchronization operations.

\punt{
\subsection{Future Work}
\label{sec:futurework}

% REMOVE, such as detecting concurrent errors in multithreaded programs. We will have to design special support in order to deterministically replay multithreaded programs. 

We plan to extend \doubletake{}'s implementation to let it work on global
variables and the stack. AddressSanitizer uses a compiler-based
approach for this purpose, which we believe would be straightforward
to integrate.

To further reduce \doubletake{}'s overhead, we plan to
replace \doubletake{}'s custom allocator with a more efficient
heap. For use-after-free detection, large objects could be allocated
directly using \texttt{mmap} and protected based on page-level
permission, rather than filling them with canaries.

\punt{
We are currently working to integrate the \doubletake{} framework
with \texttt{gdb}. Lightweight rollback and re-execution would be
useful for diagnosing application errors during a debugging
session. Additionally, a program run with \doubletake{} could
automatically begin a \texttt{gdb} session at the exact point where
the error occurred.
}

}

\section{Related Work}
\label{sec:relatedwork}

%In this section, we discuss related approaches to dynamic analysis and efficient record and replay systems.

%Dynamic analyses typically rely on one or more of the following
%approaches: dynamic instrumentation, static instrumentation, and
%interposition. We discuss prior analysis tools below, grouped by
%approach.

\textbf{Dynamic Instrumentation:} 
Numerous error detection tools use dynamic instrumentation,
including many commercial tools. Valgrind's Memcheck tool, Dr. Memory,
Purify, Intel Inspector, and Sun Discover all fall into this
category~\cite{ overflow:drmemory, overflow:purify,
overflow:inspector, overflow:valgrind, overflow:discover}. These tools
use dynamic instrumentation engines, such as Pin, Valgrind, and
DynamiRIO~\cite{DynamoRIO, Pin, overflow:valgrind}. These tools can
detect memory leaks, use-after-free errors, uninitialized reads, and
buffer overflows. Dynamic instrumentation tools are typically easy to
use because they do not require recompilation, but this ease of use
generally comes at the cost of high overhead. Programs run with
Valgrind take $20\times$ longer than usual, and Dr. Memory introduces
$10\times$ runtime overhead. \doubletake{} is \emph{significantly}
more efficient than prior dynamic instrumentation tools, with under
5\% performance overhead.

Several dynamic analysis tools leverage static analysis to reduce the
amount and thus the overhead of
instrumentation~\cite{overflow:Baggy, overflow:Mudflap, overflow:lbc, overflow:ccured, 
overflow:Insure++, AddressSanitizer}.  While these
tools generally reduce overhead over approaches based exclusively on
dynamic instrumentation, but cannot detect errors in code that was not
recompiled with this instrumentation in place (e.g., inside
libraries). In addition, our results show that AddressSanitizer (the
previous state-of-the-art, which depends on static analysis) is
considerably slower than \doubletake{}, which can also perform its
analysis on the entire program (including libraries) with no
recompilation.

\textbf{Interposition:}
\doubletake{} uses library interposition exclusively during normal
execution. More expensive instrumentation is only introduced after an
error has been detected. BoundsChecker interposes on Windows heap
library calls to detect memory leaks, use-after-free errors and buffer
overflows~\cite{BoundsChecker}. Many prior approaches use a mix of
library interposition and virtual memory techniques to detect memory
errors~\cite{duma, Sleigh, Undangle, SWAT, Dlmalloc, GuardMalloc,
exterminator, Hound, electricfence,
overflow:Cruiser}, though their overhead is much higher than \doubletake{}'s.
 
\textbf{Record and replay:}
Several replay-based approaches target software debugging and/or
fault tolerance~\cite{Bressoud:1995:HFT:224056.224058, OSDebug, Rx,
RecPlay, Flashback, Triage}. Flashback supports replay
debugging by employing a shadow process to checkpoint the state of an
application, and recording the results of system calls to facilitate
the replay. Triage uses replay to automate the
failure diagnosis process for crashing bugs~\cite{Triage}. Both
Flashback and Triage need custom kernel support. 
%%Aftersight and Speck use record and replay for dynamic analysis, but incur substantially higher overhead than \doubletake{}~\cite{Aftersight,Speck}.

Aftersight is the related work that is closest in spirit
to \doubletake{}~\cite{Aftersight}. 
It separates analysis from normal
execution by logging inputs to a virtual machine and exporting them to
a separate virtual machine for detailed (slow) analysis that can run
offline or concurrently with application execution. 
Aftersight monitors applications
running in a virtual machine, which adds some amount of
workload-dependent overhead. VM-based recording alone adds additional
runtime overhead, an average of 5\% on the SPEC CPU2006
benchmarks. Aftersight's dynamic analyses are offloaded to unused
processors, which may not be available in some deployments. Unlike
Aftersight, \DoubleTake{} does not require the use of a virtual
machine, does not rely on additional processors for dynamic analyses,
and incurs lower average overhead.

Speck is another replay-based system focused on security checking,
including taint analysis and virus scanning~\cite{Speck}. Security
checks generally require applications to halt immediately upon
detecting an error, functionality that \doubletake{} by design does
not provide. Other systems have focused on reducing the performance
overhead of recording~\cite{ODR, Respec, PRES, DoublePlay}.
%\doubletake{} will borrow some of their ideas and
%provide the deterministic replay for multithreaded programs in the
%future.
%\todo{However, don't they do deterministic replay for MT programs?}

%\subsection{Evidence-based Approaches}
%DataCollider~\cite{Erickson:2010:EDD:1924943.1924954} detects data races using the evidence-based approach.
% O'Callahan's chronicle and chronomancer
%http://robert.ocallahan.org/2007/08/announcing-chronomancer\_21.html

% eliminate global lock
% possibly adopting a page-ownership protocol as used by ...

%\newpage

\section{Conclusion}
\label{sec:conclusion}

This paper introduces \emph{evidence-based dynamic analysis}, a new
lightweight dynamic analysis technique. Evidence-based dynamic
analysis works for errors that naturally leave evidence of their
occurrence, or can be forced to do so. These errors include key
problems for C and C++ programs: buffer overflows, dangling-pointer
errors, and memory leaks. Evidence-based dynamic analysis is fast
because it lets the application run at full speed until an error is
detected; execution is then rolled back and replayed with
instrumentation at the point where the evidence was found, pinpointing
the error. We present \doubletake{}, an evidence-based dynamic
analysis framework, and implement these analyses using it. The
resulting analyses are the fastest to date, imposing on average under
5\% overhead. These results demonstrate the effectiveness and
efficiency of this approach, which promises to speed testing and dramatically
increase the reach of dynamic analysis by extending it to deployed settings.
\doubletake{} is available for download at \url{http://github.com/plasma-umass/DoubleTake}.
% \url{http://github.com/removed-for-double-blind}.

\section{Acknowledgements}

This material is based upon work supported by the National Science
Foundation under Grant No. CNS-1525888. Charlie Curtsinger was
supported by a Google PhD Research Fellowship. We thank Kostya
Serebryany for his assistance with our evaluation of AddressSanitizer,
and Qiang Zeng, Dinghao Wu and Peng Liu for providing the test cases
used in their Cruiser paper.  We also thank Scott Kaplan for his
suggestions and comments during the development of
\doubletake{}. Finally, we thank John Vilk for his invaluable
suggestions that helped improve this paper.

%\newpage

{
\bibliographystyle{abbrv}
\bibliography{refs}
}

\end{document}